\documentstyle[12pt,epsf]{article}

\def\npb#1#2#3{    {\it Nucl. Phys. }{\bf B #1} (19#2) #3}
\def\plb#1#2#3{    {\it Phys. Lett. }{\bf B #1} (19#2) #3}
\def\prd#1#2#3{    {\it Phys. Rev. }{\bf D #1} (19#2) #3}
\def\prep#1#2#3{   {\it Phys. Rep. }{\bf #1} (19#2) #3}
\def\prl#1#2#3{    {\it Phys. Rev. Lett. }{\bf #1} (19#2) #3}

\def\zpc#1#2#3{    {\it Zeit. f\"ur Physik }{\bf C #1} (19#2) #3}

\def\Im{\mathop{\mbox{Im}}}
\def\Re{\mathop{\mbox{Re}}}

\def\ss{\scriptsize}

\newcommand{\beqn}{\begin{eqnarray}}
\newcommand{\beq}{\begin{equation}}
\newcommand{\eeqn}{\end{eqnarray}}
\newcommand{\eeq}{\end{equation}}
\newcommand{\nn}{\nonumber}

\begin{document}
\begin{titlepage}
\begin{flushright}
{
TUM-HEP-303/97 \\
November 1997
}
\end{flushright}
\vskip 0.6cm
\centerline {\Large{\bf Two Lectures on FCNC and CP Violation}}
\vskip 0.2cm
\centerline {\Large{\bf  in Supersymmetry$^{*}$}} \normalsize
 
\vskip 0.6cm
\centerline {A. Masiero}
\centerline {\it SISSA, Via Beirut 2-4, 34013 Trieste, Italy and}
\centerline {\it INFN, sez. di Perugia, Via Pascoli, I-06100 Perugia, Italy}
\vskip 0.2cm
\centerline{\it and}
\vskip 0.2cm
\centerline {L. Silvestrini}
\centerline {\it Technische Universit\"{a}t M\"{u}nchen, Physik Department,}
\centerline {\it D-85748 Garching, Germany}

\vskip .65cm
 
\begin{abstract}
These two lectures constitute a reappraisal and an update of the status of the
Flavour Changing Neutral Current (FCNC) and CP violation issues in
supersymmetry. The first lecture discusses these points in the framework
of the Minimal Supersymmetric Standard Model (MSSM), while the second one
provides an analysis in a generic low-energy supersymmetric extension of
the Standard Model. The goal of these lectures is twofold: on one hand we
present a qualitative and quantitative discussion of the threat that FCNC
and CP violation represent on supersymmetry model building; on the other
hand, we point out how precious FCNC and CP violation may be in obtaining
some signals of the presence of supersymmetry at low energy in the years
that separate us from the advent of LHC physics. In particular, concerning
this latter point, we emphasize and thoroughly analyze the role of
experimental searches for rare $B$ decays and CP violation in $B$ physics.  
\end{abstract}
\vskip 0.65cm
\centerline{\small{$^{*}$ Lectures given by A. Masiero at the International
  School on Subnuclear Physics, }} 
\centerline{\small{35th Course: ``Highlights: 50 Years
Later", Erice, Italy, 26 August-4 September 1997}} 
\centerline{\small{ and at the
  International School of Physics Enrico Fermi,  Course CXXVVII:}}
\centerline{\small{  ``Heavy
flavour physics: a probe of Nature's grand design", Varenna, Italy, 8-18
July 1997.}}

\vfill

\end{titlepage}

\newpage
\section{Introduction}
\label{sec:intro}
In spite of the extraordinary success of the Standard Model (SM) in
accounting for all the existing experimental data, we have several
well-motivated theoretical reasons to expect new physics beyond it. In
this view the SM may be regarded as the low-energy limit of this more
fundamental underlying theory. Indeed, it is likely that we have a ``tower"
of underlying theories which show up at different energy scales. 

If we accept the above point of view we may try to find signals of new
physics considering the SM as a truncation to renormalizable operators of
an effective low-energy theory which respects the  $SU(3)\otimes SU(2) \otimes
U(1)$
symmetry  and whose fields are just those of the SM. The renormalizable
(i.e. of canonical dimension less or equal to four) operators giving rise
to the SM enjoy three crucial properties which have no reason to be shared
by generic operators of dimension larger than four. They are the
conservation (at any order in perturbation theory) of Baryon (B) and
Lepton (L) numbers and an adequate suppression of Flavour Changing Neutral
Current (FCNC) processes through the GIM mechanism.  

Now consider the new physics (directly above the SM in the
``tower" of new physics theories) to have a typical energy scale $\Lambda$.
In the low-energy effective Lagrangian such scale appears with a positive
power only in the quadratic scalar term (scalar mass) and in the dimension
zero operator which can be considered a cosmological constant. Notice that
$\Lambda$ cannot appear in dimension three operators related to fermion
masses because chirality forbids direct fermion mass terms in the
Lagrangian. Then in all operators of dimension larger than four $\Lambda$
will show up in the denominator with powers increasing with the dimension
of the corresponding operator. 

The crucial question that all of us, theorists and experimentalists, ask
ourselves is: where is $\Lambda$? Namely is it close to the electroweak
scale (i.e. not much above $100$ GeV) or is $\Lambda$ of the order of the
grand unification scale or the Planck scale? B- and L-violating processes
and FCNC phenomena represent a potentially interesting clue to answer this
fundamental question.

Take $\Lambda$ to be close to the electroweak scale. Then we may expect
non-renormalizable operators with B, L and flavour violations not to be
largely suppressed by the presence of powers of $\Lambda$ in the
denominator. Actually this constitutes in general a formidable challenge
for any model builder who wants to envisage new physics close to $M_W$.
Theories with dynamical breaking of the electroweak symmetry
(technicolour) and low-energy supersymmetry constitute examples of new
physics with a ``small" $\Lambda$. In these lectures we will see that the
above general considerations on potentially large B, L and flavour violations
apply to the SUSY case (it is well-known that FCNC represent a major
problem also in technicolour schemes).

Alternatively, given the abovementioned potential danger of having a small
$\Lambda$, one may feel it safer to send $\Lambda$ to superlarge values.
Apart from kind of ``philosophical" objections related to the unprecedented  
gap of many orders of magnitude without any new physics, the above
discussion points out a typical problem of this approach. Since the
quadratic scalar terms have a coefficient in front scaling with
$\Lambda^2$ we expect all scalar masses to be of the order of the
superlarge scale $\Lambda$. This is the gauge hierarchy problem and it
constitutes the main (if not only) reason to believe that SUSY should be a
low-energy symmetry.

Notice that the fact that SUSY should be a fundamental symmetry of Nature
(something of which we have little doubt given the ``beauty" of this
symmetry) 
does not imply by any means that SUSY should be a low-energy symmetry,
namely that it should hold unbroken down to the electroweak scale.
SUSY may well be present in
Nature but be broken at some very large scale (Planck scale or string
compactification scale). In that case SUSY would be of no use in tackling
the gauge hierarchy problem and its phenomenological relevance would be
practically zero. On the other hand if we invoke SUSY to tame the growth
of the scalar mass terms with the scale $\Lambda$, then we are forced to
take the view that SUSY should hold as a good symmetry down to a scale
$\Lambda$ close to the electroweak scale. Then B, L and FCNC may be useful
for us to shed some light on the properties of the underlying theory from
which the low-energy SUSY Lagrangian resulted. Let us add that there is an
independent argument in favour of this view that SUSY should be a
low-energy symmetry. The presence of SUSY partners at low energy creates
the conditions to have a correct unification of the strong and electroweak
interactions. If they were at $M_{\rm Planck}$ and the SM were all the physics
up to superlarge scales, the program of achieving such a unification
would largely fail, unless one complicates the non-SUSY GUT scheme with a
large number of Higgs representations and/or a breaking chain with
intermediate mass scales is invoked.

In these lectures we will tackle some of the major aspects of the issue
of FCNC and CP violation in SUSY theories. First of all we will discuss
whether we expect
SUSY to be actually related to the flavour problem. Then we will proceed
to analyze the different status of SUSY in relation to the FCNC problem
according to the class of SUSY models one considers. This discussion is
not a mere academic exercise, but it has profound phenomenological
implications. When you read or hear sentences starting with  ``SUSY
predicts that\dots" or ``this result would rule out SUSY\dots"  it should be
kept in mind that we do not have a low-energy SUSY theory like we have the
SM, but rather we have classes of models which differ in the content of
superfields, in their couplings, in the nature of the SUSY breaking terms,
etc. At the end of these lectures we hope that the following message may
emerge: while, undoubtedly, FCNC represents a challenge for SUSY, at the
same time it can be seen as one of the major hopes that we have now (where
now may actually mean any time before LHC!) to have some signal of the
presence of low-energy SUSY.

\section{FCNC and SUSY}
\label{sec:FCNC}

The generation of fermion masses and mixings (``flavour problem") gives 
rise to a first and important distinction among theories of new physics 
beyond the electroweak standard model. 

One may conceive a 
kind of new physics which is completely ``flavour blind", i.e. new 
interactions which have nothing to do with the flavour structure. To 
provide an example of such a situation, consider a scheme where flavour 
arises at a very large scale (for instance the Planck mass) while new 
physics is represented by a supersymmetric extension of the SM 
with supersymmetry broken at a much lower scale and with the SUSY 
breaking transmitted to the observable sector by flavour-blind gauge 
interactions. In this case one may think that the new physics does not 
cause any major change to the original flavour structure of the SM, 
namely that the pattern of fermion masses and mixings is compatible with 
the numerous and demanding tests of flavour changing neutral currents.

Alternatively, one can conceive a new physics which is entangled 
with the flavour problem. As an example consider a technicolour scheme 
where fermion masses and mixings arise through the exchange of new gauge 
bosons which mix together ordinary and technifermions. Here we expect 
(correctly enough) new physics to have potential problems in 
accommodating the usual fermion spectrum with the adequate suppression 
of FCNC. As another example of new physics which is not flavour blind, 
take a more conventional SUSY model which is derived from a 
spontaneously broken N=1 supergravity and where the SUSY breaking 
information is conveyed to the ordinary sector of the theory through 
gravitational interactions. In this case we may expect that the scale at 
which flavour arises and the scale of SUSY breaking are not so different 
and possibly the mechanism itself of SUSY breaking and transmission is 
flavour-dependent. Under these circumstances we may expect 
a potential flavour problem to arise, namely that SUSY contributions to 
FCNC processes are too large.   

The potentiality of probing SUSY in FCNC phenomena was readily realized when 
the era of SUSY  phenomenology started in the early 80's \cite{susy2}.
 In particular, the 
major implication that the scalar partners of quarks of the same electric 
charge but belonging to different generations had to share a remarkably high 
mass degeneracy was emphasized. 

Throughout the large amount of work  in this last decade it became clearer 
and clearer that generically talking of the implications of low-energy SUSY on 
FCNC may be rather misleading. We have a minimal SUSY extension of the SM, the 
so-called Minimal Supersymmetric Standard Model (MSSM) \cite{susy1}, 
where the FCNC 
contributions can be computed in terms of a very limited set of unknown 
new SUSY parameters. Remarkably enough, this minimal model succeeds to
pass all  
the set of FCNC tests unscathed. To be sure, it is possible to severely 
constrain the SUSY parameter space, for instance using $b 
\to s \gamma$, in a way which is complementary to what is achieved by direct 
SUSY searches at colliders.

However,  the MSSM is by no means equivalent to low-energy SUSY. A first 
sharp distinction concerns the mechanism of SUSY breaking and 
transmission to the observable sector which is chosen. As we mentioned 
above, in models with gauge-mediated SUSY breaking (GMSB models
\cite{GMSB1}-\cite{GMSB3})
it may be possible to avoid the 
FCNC threat ``ab initio" (notice that this is not an automatic feature of 
this class of models, but it depends on the specific choice of the 
sector which transmits the SUSY breaking information, the so-called 
messenger sector). The other more ``canonical" class of SUSY theories 
that was mentioned above has gravitational messengers and a very large 
scale at which SUSY breaking occurs. In this talk we will focus only on 
this class of gravity-mediated SUSY breaking models. Even sticking to 
this more limited choice we have a variety of options with very 
different implications for the flavour problem. 

First, there exists an interesting large class of SUSY realizations 
where the customary R-parity (which is invoked to suppress proton decay) 
is replaced by 
other discrete symmetries which allow either baryon or lepton violating terms 
in the superpotential. But, even sticking to the more orthodox view of 
imposing R-parity, we are still left with a large variety of extensions of the 
MSSM at low energy. The point is that low-energy SUSY ``feels" the new physics 
at the superlarge scale at which supergravity  (i.e., local supersymmetry) 
broke down. In this last couple of years we have witnessed an increasing 
interest in supergravity realizations without the so-called flavour 
universality of the terms which break SUSY explicitly. Another class of 
low-energy SUSY realizations which differ from the MSSM in the FCNC sector 
is obtained from SUSY-GUT's. The interactions involving superheavy particles 
in the energy range between the GUT and the Planck scale bear important 
implications for the amount and kind of FCNC that we expect at low energy.

\section{FCNC in SUSY without R-Parity}
\label{sec:Rbroken}

It is well known that in the SM case the imposition of gauge symmetry and the 
usual gauge assignment of the 15 elementary fermions of each family lead to 
the automatic conservation of baryon and lepton numbers (this is true
at any order in perturbation theory).

On the contrary, imposing in addition to the usual $SU(3)\otimes SU(2) \otimes 
U(1)$ gauge symmetry an N=1 global SUSY does not prevent the appearance of 
terms which explicitly break B or L \cite{weinb}.
 Indeed, the superpotential reads:
\beqn
W&=&h^U Q H_{U}u^c + h^D Q H_{D} d^c + h^L L H_D e^c + \mu H_U H_D \nn \\
&+& \mu^\prime H_{U} L + \lambda^{\prime \prime}_{ijk}u^c_{i}d^c_{j}d_{k}^c +
\lambda^{\prime}_{ijk}Q_{i}L_{j}d_{k}^c + \lambda_{ijk}L_{i}L_{j}e_{k}^c \, ,
\label{superp}
\eeqn
where the chiral matter superfields $Q$, $u^c$, $d^c$, $L$, $e^c$, $H_{U}$ and 
$H_{D}$ transform under the above gauge symmetry as:
\beqn
&\,&Q\equiv (3,2,1/6); \qquad u^c\equiv (\bar{3},1,-2/3);\qquad d^c\equiv
(\bar{3},1,1/3);\\
&\,& L\equiv (1,2,-1/2); \; \; e^c \equiv (1,1,1); \;\; H_{U}\equiv 
(1,2,1/2); \;\; H_{D}\equiv (1,2,-1/2). \nn
\label{qnumbers}
\eeqn
The couplings $h^U$, $h^D$, $h^L$ are $3\times 3$ matrices in the generation 
space; $i$, $j$ and $k$ are generation indices. Using the product of 
$\lambda^\prime$ and $\lambda^{\prime \prime}$ couplings it is immediate to 
construct four-fermion operators leading to proton decay through the exchange 
of a squark. Even if one allows for the existence of $\lambda^\prime$ and 
$\lambda^{\prime \prime}$ couplings only involving the heaviest generation, 
one can show that the bound on the product $\lambda^\prime \times 
\lambda^{\prime \prime}$ of these couplings is very severe (of $O(10^{-7})$)
\cite{smirnov}.

A solution is that there exists a discrete symmetry, B-parity \cite{b}, 
which forbids 
the B violating terms in eq.~(\ref{superp}) which are proportional to 
$\lambda^{\prime \prime}$. In that case it is still possible to produce 
sizeable effects in FC $B$ decays. For instance, using the product of 
$\lambda^\prime_{3jk}\lambda_{ljl^c}$ one can obtain $b \to s \,(d) + l l^c$ 
taking $k=2 \, (1)$ and through the mediation of the sneutrino of the $j$-th 
generation. Two general features of these R-parity violating contributions 
are:
\begin{enumerate}
\item  complete loss of any correlation to the CKM elements. For instance, in 
the above example, the couplings $\lambda^\prime$ and $\lambda$ have nothing 
to do with the usual angles $V_{tb}$ and $V_{ts}$ which appear in $b \to s l^+ 
l^-$ in the SM;
\item  loss of correlation among different FCNC processes which are 
tightly correlated in the SM. For instance, in our example $b \to d l^+ l^-$ 
would depend on $\lambda^\prime$ and $\lambda$ parameters which are different 
from those appearing in $B_{d}-\bar{B}_{d}$ mixing.
\end{enumerate}

In this context it is difficult to make predictions given the arbitrariness of 
the large number of $\lambda$ and $\lambda^\prime$ parameters. There exist 
bounds on each individual coupling (i.e. assuming all the other L violating 
couplings are zero) \cite{barger}.
With some exception, they are not very stringent for the third generation 
(generally of $O(10^{-1})$), 
hence allowing for conspicuous effects. Indeed, one 
may think of using the experimental bounds on rare $B$ decays to put severe 
bounds on products of L violating couplings.

Obviously, the most practical way of avoiding any threat of B and L violating 
operators is to forbid \underline{all} such terms in eq.~(\ref{superp}). This 
is achieved by imposing the usual R matter parity. This quantum number reads 
$+1$ over every ordinary particle and $-1$ over SUSY partners. We now turn to 
FCNC in the framework of low-energy SUSY with R parity.

\section{FCNC in SUSY with R Parity - MSSM Framework}
\label{sec:MSSM}

Even when R parity is imposed the FCNC challenge is not over. It is true
that in this case, analogously to what happens in the SM, no tree
level FCNC contributions arise. However, it is well-known that this is a
necessary but not sufficient condition to consider the FCNC problem
overcome. The loop contributions to FCNC in the SM exhibit the presence of
the GIM mechanism and we have to make sure that in the SUSY case with R
parity some analog of the GIM mechanism is active. 

To give a qualitative idea of what we mean by an effective super-GIM
mechanism, let us consider the following simplified situation where the
main features emerge clearly. Consider the SM box diagram responsible for
the $K^0 - \bar{K}^0$ mixing and take only two generations, i.e. only the up
and charm quarks run in the loop. In this case the GIM mechanism yields a
suppression factor of $O((m_c^2 - m_u^2)/M_W^2)$. If we replace the W
boson and the up quarks in the loop with their SUSY partners and we take,
for simplicity, all SUSY masses of the same order, we obtain a
super-GIM factor which looks like the GIM one with the masses of the
superparticles instead of those of the corresponding particles. The
problem is that the up  and charm squarks have masses 
which are much larger
than those of the corresponding quarks. Hence the super-GIM factor tends to
be of $O(1)$ instead of being $O(10^{-3})$ as it is in the SM case. To
obtain this small number we would need a high degeneracy between the mass of
the charm and up squarks. It is difficult to think that such a degeneracy
may be accidental. After all, since we invoked SUSY for a naturalness
problem (the gauge hierarchy issue), we should avoid invoking
a fine-tuning to solve its problems! Then one can turn to some symmetry
reason. For instance, just sticking to this simple example that we are
considering, one may think that the main bulk of the charm and up squark
masses is the same, i.e. the mechanism of SUSY breaking should have some
universality in providing the mass 
to these two squarks with the same
electric charge. Another possibility one may envisage is that the masses
of the squarks are quite high, say above few TeV's. Then even if they are
not so degenerate in mass, the overall factor in front of the four-fermion
operator responsible for the kaon mixing becomes smaller and smaller (it
decreases quadratically with the mass of the squarks) and, consequently, one
can respect the observational result. We see from this simple example
that the issue of FCNC may be closely linked to the crucial problem of the
way we break SUSY.

We now turn to some more quantitative considerations. We start by
discussing the different degree of concern that FCNC originate according
to the specific low-energy SUSY realization one has in mind. In this section
we will consider FCNC in the MSSM realizations. In Sect. \ref{sec:MSSMCP}
we will  deal with CP-violating FCNC phenomena in the
same context. 
 After discussing 
these aspects in the MSSM we will 
provide bounds from FCNC and CP violation in a generic SUSY extension of the
SM (Sect. \ref{sec:genFCNC} and \ref{sec:genCP}).

Obviously the reference frame for any discussion in a specific SUSY scheme
is the MSSM. Although
the name seems to indicate a well-defined particle model, actually  
MSSM denotes at least two quite different classes of low-energy SUSY models.
In its most restrictive meaning it denotes the minimal SUSY extension of
the  
SM (i.e. with the smallest needed number of superfields) with R-parity,
radiative breaking of the electroweak symmetry, universality of the soft
breaking
terms and simplifying relations at the GUT scale among SUSY parameters. In
this ``minimal" version the MSSM exhibits only four free parameters in
addition to those of the SM. Moreover, some authors impose specific relations
between the two parameters $A$ and $B$ that appear in the trilinear and
bilinear scalar terms of the soft breaking sector further reducing the number
of SUSY free parameters to three. Then, all SUSY masses are just function of
these few independent parameters and, hence, many relations among them
exist. 
Obviously this very minimal version of the MSSM can be very predictive. The
most powerful constraint on this minimal model in the FCNC context comes from
$b \to s \gamma$.

In SUSY there are five classes of one-loop diagrams which contribute
to FCNC $B$ decays. They are distinguished according to the virtual
particles running in the loop: W and up-quarks, charged Higgs and
up-quarks, charginos and up-squarks, neutralinos and down-squarks,
gluinos and down-squarks. It turns out that, at least in this
``minimal" version of the MSSM, the charged Higgs and chargino
exchanges yield the dominant SUSY contributions. We will deal more on
this point in the next section.

 As for $b \to s
\gamma$ the situation can be summarized as follows. The CLEO measurement 
yields BR$(B \to X_{s}\gamma)=(2.32 \pm 0.67)\times 10^{-4}$ \cite{cleo}.
On the
theoretical side we have just witnessed a major breakthrough
with the
computation of the next-to-leading logarithmic result for the BR. This has   
been
achieved thanks to the calculation of the $O(\alpha_{s})$ matrix elements
\cite{greub} and
of the next-to-leading order Wilson coefficients at $\mu \simeq m_{b}$
\cite{misiak}. The present theoretical result,
 BR$(B \to X_{s} \gamma)=(3.28 \pm 0.33)
\times 10^{-4}$ \cite{misiak}, exhibits an impressive improvement on the
size of
the error. A substantial improvement also on the experimental error is
foreseen for the near future. Hence $b \to s \gamma$ is going to constitute
the most relevant place in FCNC $B$ physics to constrain SUSY at least before
the advent of $B$ factories. So far this process has helped in ruling out
regions of the SUSY parameter space which are even larger than those excluded
by LEP I and it is certainly going to be complementary to what LEP II is
expected to do in probing the SUSY parameter space. After the detailed
analysis in 1991 \cite{bertol}
for small values of $\tan \beta$, there have been recent
analyses \cite{barb}
  covering the entire range of $\tan \beta$ and including also other
technical improvements (for instance radiative corrections in the Higgs
potential). It has been shown \cite{vissani}
that the exclusion plots are very sensitive also
to the relation one chooses between $A$ and $B$. It should be kept in mind that
the ``traditional" relation $B=A-1$ holds true only in some simplified version
of the MSSM. A full discussion is beyond the scope of these lectures and so we
refer
the interested readers to the vast literature which exists on the subject.

The constraint on the SUSY parameter space of the ``minimal" version of the 
MSSM greatly affects also the potential departures of this model from the SM
expectation for $b \to s l^+ l^-$. The present limits on the exclusive
channels BR$(B^{0} \to K^{*0} e^{+} e^{-})$
 and BR$(
B^{0} \to K^{*0} \mu^{+} \mu^{-})$
are within an
order of magnitude of the SM predictions. On the theoretical side, it has
been estimated that the evaluation of $\Gamma (B\to X_{s}l^{+} l^{-})$ in the
SM is going to be affected by an error which cannot be reduced to less than
$10-20 \%$ due to uncertainties in quark masses and interference effects from
excited charmonium states \cite{ligeti}.
It turns out that, keeping into account the bound
on $b \to s \gamma$, in the MSSM with universal soft breaking terms a $20 \%$
departure from the SM expected BR is kind of largest possible value one can
obtain \cite{cho}.
Hence the chances to observe a meaningful deviation in this case are
quite slim. However, it has been stressed that in view of the fact that three
Wilson
coefficients play a relevant role in the effective low-energy Hamiltonian 
involved in $b \to s \gamma$ and $b \to s l^{+} l^{-}$, a third observable in
addition to BR$(b \to s \gamma)$ and BR$(b \to s l^{+}l^{-})$ is needed.  
This has been identified in some asymmetry of the emitted leptons (see
 refs.~\cite{cho,agm} for two different choices of such asymmetry).
This quantity, even
in the ``minimal" MSSM, may undergo a conspicuous deviation from its SM
expectation and, hence, hopes of some manifestation of SUSY, even in this
minimal realization, in $b \to s l^{+} l^{-}$ are still present.

Finally, also for the $B_{d}-\bar{B}_{d}$ mixing, in the above-mentioned   
analysis of rare $B$ physics in the MSSM with universal soft breaking terms
\cite{bertol}
it was emphasized that, at least in the low $\tan \beta$ regime, one cannot
expect an enhancement larger than $20\%-30\%$ over the SM prediction
(see also ref.~\cite{kurimoto}). Moreover
it was shown that $x_{s}/x_{d}$ is expected to be the same as in the SM.

It should be kept in mind that the above stringent results strictly depend not 
only on the minimality of the model in terms of the superfields that are   
introduced, but also on the ``boundary" conditions that are chosen.
All the low-energy SUSY masses are computed in terms of the $M_{Pl}$ four
SUSY parameters through the RGE evolution. If one relaxes this tight
constraint on the relation of the low-energy quantities and treats the masses
of the SUSY particles as independent parameters, then much more freedom is
gained. This holds true even if flavour universality is enforced. For
instance,
 BR$(b \to s \gamma
)$ and $\Delta m_{B_{d}}$ may vary a lot from the SM expectation, in
particular in regions of moderate SUSY masses \cite{brignole} (the most
interesting case, i.e. small chargino and stop masses, will be briefly
dealt with in next section).

Moreover, flavour universality is by no means a prediction of low-energy SUSY.
The absence of flavour universality of soft-breaking terms may result from 
radiative effects at the GUT scale or from effective supergravities derived
from string theory. For instance, even starting with an exact universality
of the soft breaking terms at the Planck scale, in a SUSY GUT scheme one
has to consider the running from this latter scale and the GUT scale. Due
to the large value of the top Yukawa coupling and to the fact that quarks
and lepton superfields are in common GUT multiplets, we may expect the tau
slepton mass to be conspicuously different from that of the first two
generation sleptons at the end of this RG running. This lack of
universality at the GUT scale may lead to large violations of flavour
lepton number yielding, for instance, $\mu \to e  \gamma$ at a rate in the
ball park of observability \cite{strumia}. In the non-universal case,
 BR$(b \to s l^{+} l^{-})$
is strongly affected by this larger freedom in the parameter space. There are
points of this parameter space where the non-resonant BR$(B \to X_{s} e^{+}
e^{-})$ and BR$(B \to X_{s} \mu^{+}\mu^{-})$  are enhanced by up to $90 \%$ and
$110 \%$ while still respecting the constraint coming from $b \to s \gamma$
\cite{cho}.

\section{CP Violation in the MSSM}
\label{sec:MSSMCP}

CP violation has major potentialities to exhibit manifestations of new physics
beyond the standard model.
Indeed, it is quite a general feature that new physics possesses
new CP violating phases in addition to the
Cabibbo-Kobayashi-Maskawa (CKM) phase $\left(\delta_{\rm CKM}\right)$
or, even in those cases where this does not occur, $\delta_{\rm CKM}$
shows up in interactions of the new particles, hence with potential departures
from the SM expectations. Moreover, although the SM is able to account for the
observed CP violation in the kaon system, we cannot say that we have tested so
far the SM predictions for CP violation. The detection of CP violation in $B$
physics will constitute a crucial test of the standard CKM picture within the
SM. Again, on general grounds, we expect new physics to provide departures from
the SM CKM scenario for CP violation in $B$ physics. A final remark on reasons
that make us optimistic in having new physics playing a major role in CP
violation concerns the matter-antimatter asymmetry in the universe. Starting
from a baryon-antibaryon symmetric universe, the SM is unable to account for
the observed baryon asymmetry. The presence of new CP-violating contributions
when one goes beyond the SM looks crucial to produce an efficient mechanism for
the generation of a satisfactory $\Delta$B asymmetry.

The above considerations apply well to the new physics represented by
low-energy supersymmetric extensions of the SM. Indeed, as we will see below,
supersymmetry introduces CP violating phases in addition to
$\delta_{\rm CKM}$ and, even if one envisages particular situations
where such extra-phases vanish, the phase $\delta_{\rm CKM}$ itself
leads to new CP-violating contributions in processes where SUSY particles are
exchanged. CP violation in $B$ decays has all potentialities to exhibit
departures from the SM CKM picture in low-energy SUSY extensions, although, as
we will discuss, the detectability of such deviations strongly depends on the
regions of the SUSY parameter space under consideration.

In this section we will deal with CP violation in the context of the MSSM.
In Sec. \ref{sec:genCP} we will discuss the CP issue in a
model-independent approach. 
For recent reviews on CP violation in SUSY see \cite{rattazzi}.

In the MSSM two new ``genuine" SUSY CP-violating phases are present. They
originate from the SUSY parameters $\mu$, $M$, $A$ and $B$. The first of these
parameters is the dimensionful coefficient of the $H_u H_d$ term of the
superpotential. The remaining three parameters are present in the sector that
softly breaks the N=1 global SUSY. $M$ denotes the common value of the gaugino
masses, $A$ is the trilinear scalar coupling, while $B$ denotes the bilinear
scalar coupling. In our notation all these three parameters are
dimensionful. The simplest way to see which combinations of the phases of these
four parameters are physical \cite{Dugan} is to notice that for vanishing
values of $\mu$,  $M$, $A$ and $B$ the theory possesses two additional
symmetries \cite{Dimopoulos}. Indeed, letting $B$ and $\mu$ vanish, a $U(1)$
Peccei-Quinn symmetry originates, which in particular rotates $H_u$ and $H_d$.
If $M$, $A$ and $B$ are set to zero, the Lagrangian acquires a continuous
$U(1)$ $R$ symmetry. Then we can consider  $\mu$,  $M$, $A$ and $B$ as spurions
which break the $U(1)_{PQ}$ and $U(1)_R$ symmetries. In this way the question
concerning the number and nature of the meaningful phases translates into the
problem of finding the independent combinations of the four parameters which
are invariant under $U(1)_{PQ}$ and $U(1)_R$ and determining their independent
phases. There are three such independent combinations, but only two of their
phases are independent. We use here the commonly adopted choice:
\begin{equation}
  \label{MSSMphases}
  \Phi_A = {\rm arg}\left( A^* M\right), \qquad
  \Phi_B = {\rm arg}\left( B^* M\right).
\end{equation}
The main constraints on $\Phi_A$ and $\Phi_B$ come from their contribution to
the electric dipole moments of the neutron and of the electron. For instance,
the effect of $\Phi_A$ and $\Phi_B$ on the electric and chromoelectric dipole
moments of the light quarks ($u$, $d$, $s$) lead to a contribution to
$d^e_N$ of 
order \cite{EDMN}
\begin{equation}
  \label{EDMNMSSM}
  d^e_N \sim 2 \left( \frac{100 {\rm GeV}}{\tilde{m}}\right)^2 \sin \Phi_{A,B}
  \times 10^{-23} {\rm e\, cm},
\end{equation}
where $\tilde{m}$ here denotes a common mass for squarks and gluinos. The
present experimental bound, $d^e_N < 1.1 \time 10^{-25}$ e cm, implies that
$\Phi_{A,B}$ should be $<10^{-2}$, unless one pushes SUSY masses up to O(1
TeV). A possible caveat to such an argument calling for a fine-tuning of
$\Phi_{A,B}$ is that uncertainties in the estimate of the hadronic matrix
elements could relax the severe bound in eq.~(\ref{EDMNMSSM}) \cite{Ellis}.

In view of the previous considerations most authors dealing with the MSSM
prefer to simply put $\Phi_A$ and $\Phi_B$ equal to zero. Actually, one may
argue in favour of this choice by considering the soft breaking sector of the
MSSM as resulting from SUSY breaking mechanisms which force $\Phi_A$ and
$\Phi_B$ to vanish. For instance, it is conceivable that both $A$ and $M$
originate from one same source of $U(1)_R$ breaking. Since $\Phi_A$ ``measures"
the relative phase of $A$ and $M$, in this case it would ``naturally"vanish. In
some specific models it has been shown \cite{Dine} that through an analogous
mechanism also $\Phi_B$ may vanish.

If $\Phi_A=\Phi_B=0$, then the novelty of SUSY in CP violating contributions
merely arises from the presence of the CKM phase in loops where SUSY particles
run \cite{CPSUSY}. The crucial point is that the usual GIM suppression, which
plays a major role in evaluating $\varepsilon$ and $\varepsilon^\prime$ in the
SM, in the MSSM case is replaced by a super-GIM cancellation which has the same
``power" of suppression as the original GIM (see previous section). Again
also in
the MSSM as it is the case in the SM, the smallness of $\varepsilon$ and
$\varepsilon^\prime$ is guaranteed not by the smallness of
$\delta_{\rm CKM}$, but
rather by the small CKM angles and/or small Yukawa couplings. By the same
token, we do not expect any significant departure of the MSSM from the SM
predictions also concerning CP violation in $B$ physics. As a matter of fact,
given the large lower bounds on squark and gluino masses, one expects
relatively tiny contributions of the SUSY loops in $\varepsilon$ or
$\varepsilon^\prime$ in comparison with the normal $W$ loops of the SM. Let us
be more detailed on this point.

In the MSSM the gluino exchange contribution
to FCNC is subleading with respect to chargino ($\chi^\pm$) and charged
Higgs ($H^\pm$) exchanges. Hence when dealing with CP violating FCNC
processes in the MSSM with $\Phi_A=\Phi_B=0$ one can confine the analysis
 to $\chi^\pm$
and $H^\pm$ loops. If one takes all squarks to be degenerate in mass and
heavier than $\sim 200$ GeV, then $\chi^\pm-\tilde q$ loops are obviously
severely penalized with respect to the SM $W-q$ loops (remember that at the
vertices the same CKM angles occur in both cases).

The only chance for the MSSM to produce some sizeable departure from the SM
situation in CP violation is in the particular region of the parameter space
where one has light $\tilde q$, $\chi^\pm$ and/or $H^\pm$. The best
candidate (indeed the only one unless 
$\tan \beta \sim m_t/m_b$) for a light squark is the stop. Hence one can
ask the following question: can the MSSM present some novelties in CP-violating
phenomena when we consider $\chi^+ - \tilde t$ loops with light $ \tilde t$,
$\chi^+$ and/or $H^+$?

Several analyses in the literature tackle the above question or, to be more
precise, the more general problem of the effect of light $\tilde t$
and $\chi^+$ 
on FCNC processes \cite{refbrignole,mpr}. A first important
observation concerns the
relative sign of the $W-t$ loop with respect to the  $\chi^+ - \tilde t$ and
$H^+ - t$ contributions. As it is well known, the latter contribution always
interferes positively with the SM one. Interestingly enough, in the region of
the MSSM parameter space that we consider here, also the $\chi^+ - \tilde t$
contribution constructively interferes with the SM contribution. The second
point regards the composition of the lightest chargino, i.e. whether the
gaugino or higgsino component prevails. This is crucial since the light stop is
predominantly $\tilde t_R$ and, hence, if the lightest chargino is mainly a
wino then it couples to $\tilde t_R$ mostly through the $LR$ mixing in the stop
sector. Consequently, a suppression in the contribution to box diagrams going
as $\sin^4 \theta_{LR}$ is present ($\theta_{LR}$ denotes the mixing angle
between
the lighter and heavier stops). On the other hand, if the lightest chargino is
predominantly a higgsino (i.e. $M_2 \gg \mu$ in the chargino mass matrix), then
the $\chi^+-$lighter $\tilde t$ contribution grows. In this case contributions
$\propto \theta_{LR}$ become negligible and, moreover, it can be shown that
they are independent on the sign of $\mu$. A detailed study is provided in
reference \cite{mpr}. For instance, for $M_2/\mu=10$ they find that the
inclusion of
the SUSY contribution to the box diagrams doubles the usual SM contribution for
values of the lighter $\tilde t$ mass up to $100-120$ GeV, using $\tan \beta
=1.8$, $M_{H^+}=100$ TeV, $m_\chi=90$ GeV and the mass of the heavier $\tilde
t$ of 250 GeV. However, if $m_\chi$ is pushed up to 300 GeV, the  $\chi^+ -
\tilde t$ loop yields a contribution which is roughly 3 times less than in the
case $m_\chi=90$ GeV, hence leading to negligible departures from the SM
expectation. In the cases where the SUSY contributions are sizeable, one
obtains relevant restrictions on the $\rho$ and $\eta$ parameters of the CKM
matrix by making a fit of the parameters $A$, $\rho$ and $\eta$ of the CKM
matrix and of the total loop contribution to the experimental values of
$\varepsilon_K$ and $\Delta M_{B_d}$. For instance, in the above-mentioned
case in which the SUSY loop contribution equals the SM $W-t$ loop, hence giving
a total loop contribution which is twice as large as in the pure SM case,
combining the $\varepsilon_K$ and $\Delta M_{B_d}$ constraints leads to a
region in the $\rho-\eta$ plane with $0.15<\rho<0.40$ and $0.18<\eta<0.32$,
excluding negative values of $\rho$.

In conclusion, the situation concerning CP violation in the MSSM case with
$\Phi_A=\Phi_B=0$ and exact universality in the soft-breaking sector can be
summarized in the following way: the MSSM does not lead to any significant
deviation from the SM expectation for CP-violating phenomena as $d_N^e$,
$\varepsilon$, $\varepsilon^\prime$ and CP violation in $B$ physics; the only
exception to this statement concerns a small portion of the MSSM
parameter space 
where a very light $\tilde t$ ($m_{\tilde t} < 100$ GeV) and $\chi^+$
($m_\chi \sim 90$ GeV) are present. In this latter particular situation
sizeable SUSY contributions to $\varepsilon_K$ are possible and, consequently,
major restrictions in the $\rho-\eta$ plane can be inferred. Obviously, CP
violation in $B$ physics becomes a crucial test for this MSSM case with very
light $\tilde t$ and $\chi^+$. Interestingly enough, such low values of SUSY
masses are at the border of the detectability region at LEP II.

\section{Model-Independent Analysis of FCNC Processes in SUSY}
\label{sec:genFCNC}

Given a specific SUSY model it is in principle possible to make a full
computation of all the FCNC phenomena in that context. However, given the
variety of options for low-energy SUSY which was mentioned in the Introduction
(even confining ourselves here to models with R matter parity), it is
important to have a way to extract from the whole host of FCNC processes a set
of upper limits on quantities which can be readily computed in any chosen SUSY
frame.

The best model-independent parameterization of FCNC effects is the so-called
mass insertion approximation \cite{mins}.
It concerns the most peculiar source of FCNC SUSY contributions that do not
arise from the mere supersymmetrization of the FCNC in the SM. They originate
from the FC couplings of gluinos and neutralinos to fermions and
sfermions~\cite{FCNC}. One chooses a basis
for the fermion and sfermion states where all the couplings of these particles
to neutral gauginos are flavour diagonal, while the FC is exhibited by the
non-diagonality of the sfermion propagators. Denoting by $\Delta$ the   
off-diagonal terms in the sfermion mass matrices (i.e. the mass terms relating
sfermion of the same electric charge, but different flavour), the sfermion
propagators can be expanded as a series in terms of $\delta = \Delta/
\tilde{m}^2$
where $\tilde{m}$ is the average sfermion mass.
As long as $\Delta$ is significantly smaller than $\tilde{m}^2$,
we can just take
the first term of this expansion and, then, the experimental information
concerning FCNC and CP violating phenomena translates into upper bounds on
these $\delta$'s \cite{deltas}-\cite{GGMS}.

  Obviously the above mass insertion method presents the major advantage that
 one does not need the full diagonalization of the sfermion mass matrices to
 perform a test of the SUSY model under consideration in the FCNC sector. It is
enough to compute ratios of the off-diagonal over the diagonal entries of the
 sfermion mass matrices and compare the results with the general bounds on the
 $\delta$'s that we provide here from all available experimental information.

  There exist four different   
$\Delta$ mass insertions connecting flavours $i$ and $j$
 along a sfermion propagator: $\left(\Delta_{ij}\right)_{LL}$,
$\left(\Delta_{ij}\right)_{RR}$, $\left(\Delta_{ij}\right)_{LR}$ and
$\left(\Delta_{ij}\right)_{RL}$. The indices $L$ and $R$ refer to the
helicity of
the
fermion partners. The size of these $\Delta$'s can be quite different. For
instance, it is well known that in the MSSM case, only the $LL$ mass insertion
can change flavour, while all the other three above mass insertions are flavour
 conserving, i.e. they have $i=j$. In this case to realize a $LR$ or $RL$ 
flavour
 change one needs a double mass
insertion with the flavour changed solely in a $LL$
 mass insertion and a subsequent flavour-conserving $LR$ mass insertion.
Even worse
 is the case of a FC $RR$ transition: in the MSSM this can be accomplished only
 through a laborious set of three mass insertions, two flavour-conserving $LR$
transitions and an $LL$ FC insertion.
  Instead of the dimensional quantities $\Delta$ it is more
useful to provide bounds making use of dimensionless quantities, $\delta$,
that are obtained dividing the mass insertions by an average sfermion mass.

Let us first consider CP-conserving $\Delta F=2$ processes.
The amplitudes for gluino-mediated contributions to $\Delta F=2$ transitions
in the mass-insertion approximation have been computed in
refs.~\cite{GMS,GGMS}. Imposing that the contribution to $K-\bar K$,
$D-\bar D$ and $B_d - \bar{B}_d$ mixing
proportional to
each single $\delta$ parameter does not exceed the experimental value,
we obtain the constraints on the $\delta$'s reported in
table~\ref{reds2}, barring accidental cancellations \cite{GGMS} (for
a QCD-improved computation of the constraints coming from $K-\bar K$
mixing, see ref.~\cite{bagger}).

\begin{table}
 \begin{center}
 \begin{tabular}{||c|c|c|c||}  \hline \hline
 $x$ & $\sqrt{\left|\Re  \left(\delta^{d}_{12} \right)_{LL}^{2}\right|} $ 
 &
 $\sqrt{\left|\Re  \left(\delta^{d}_{12} \right)_{LR}^{2}\right|} $ &
 $\sqrt{\left|\Re  \left(\delta^{d}_{12} \right)_{LL}\left(\delta^{d}_{12}
 \right)_{RR}\right|} $ \\
 \hline
 $
   0.3
 $ &
 $
1.9\times 10^{-2}
 $ & $
7.9\times 10^{-3}
 $ & $
2.5\times 10^{-3}
 $ \\
 $
   1.0
 $ &
 $
4.0\times 10^{-2}
 $ & $
4.4\times 10^{-3}
 $ & $
2.8\times 10^{-3}
 $ \\
 $
   4.0
 $ &
 $
9.3\times 10^{-2}
 $ & $
5.3\times 10^{-3}
 $ & $
4.0\times 10^{-3}
 $ \\ \hline \hline
 $x$ & $\sqrt{\left|\Re  \left(\delta^{d}_{13} \right)_{LL}^{2}\right|} $ 
 &
 $\sqrt{\left|\Re  \left(\delta^{d}_{13} \right)_{LR}^{2}\right|} $ &
 $\sqrt{\left|\Re  \left(\delta^{d}_{13} \right)_{LL}\left(\delta^{d}_{13}
 \right)_{RR}\right|} $ \\
 \hline
 $
   0.3
 $ &
 $
4.6\times 10^{-2}
 $ & $
5.6\times 10^{-2}
 $ & $
1.6\times 10^{-2}
 $ \\
 $
   1.0
 $ &
 $ 
9.8\times 10^{-2}
 $ & $
3.3\times 10^{-2}
 $ & $
1.8\times 10^{-2}
 $ \\
 $
   4.0
 $ &
 $
2.3\times 10^{-1}
 $ & $
3.6\times 10^{-2}
 $ & $
2.5\times 10^{-2}
 $ \\ \hline \hline
 $x$ & $\sqrt{\left|\Re  \left(\delta^{u}_{12} \right)_{LL}^{2}\right|} $ 
 &
 $\sqrt{\left|\Re  \left(\delta^{u}_{12} \right)_{LR}^{2}\right|} $ &
 $\sqrt{\left|\Re  \left(\delta^{u}_{12} \right)_{LL}\left(\delta^{u}_{12}
 \right)_{RR}\right|} $ \\
 \hline
 $
   0.3
 $ &
 $
4.7\times 10^{-2}
 $ & $
6.3\times 10^{-2}
 $ & $
1.6\times 10^{-2}
 $ \\
 $
   1.0
 $ &
 $
1.0\times 10^{-1}
 $ & $
3.1\times 10^{-2}
 $ & $
1.7\times 10^{-2}
 $ \\
 $
   4.0
 $ &
 $
2.4\times 10^{-1}
 $ & $
3.5\times 10^{-2}
 $ & $
2.5\times 10^{-2}
 $ \\ \hline \hline
 \end{tabular}
 \caption[]{Limits on $\mbox{Re}\left(\delta_{ij}\right)_{AB}\left(
 \delta_{ij}\right)_{CD}$, with $A,B,C,D=(L,R)$, for an average squark mass
 $m_{\tilde{q}}=500\mbox{GeV}$ and for different values of 
 $x=m_{\tilde{g}}^2/m_{\tilde{q}}^2$. For different values of $m_{\tilde{q}}$, 
 the limits can be obtained multiplying the ones in the table by 
 $m_{\tilde{q}}(\mbox{GeV})/500$.}
 \label{reds2}
 \end{center}
 \end{table} 

We then consider the process $b \to s \gamma$. This decay requires a helicity
flip. In the presence of a 
$\left(\delta^d_{23}\right)_{LR}$ mass insertion we can realize this flip in
the gluino running in the loop. On the contrary, the $\left(
\delta^d_{23}\right)_{LL}$ insertion requires the helicity flip to occur in
the external $b$-quark line. Hence we expect a stronger bound on the
$\left(\delta^d_{23}\right)_{LR}$ quantity. Indeed, this is what happens:
$\left(\delta^d_{23}\right)_{LL}$ is essentially not bounded, while
$\left(\delta^d_{23}\right)_{LR}$ is limited to be $<10^{-3}-10^{-2}$
according to the average squark and gluino masses (see
table~\ref{tab:bsg}) \cite{GGMS}.
Given the upper
bound on $\left(\delta^d_{23}\right)_{LR}$ from $b \to s \gamma$, it turns out
that the quantity $x_{s}$ of the $B_{s}-\bar{B}_{s}$ mixing receives
contributions from this kind of mass insertions which are very tiny. The only
chance to obtain large values of $x_s$ is if $\left(\delta^d_{23}\right)_{LL}$
is large, say of $O(1)$. In that case $x_s$ can easily jump up to values of $O
(10^{2})$ or even larger.

Then, imposing the bounds in table~\ref{reds2}, we can obtain the
largest possible 
value for BR($b \to d \gamma$) through gluino exchange. As expected, the
$\left( \delta^{d}_{13}\right)_{LL}$ insertion leads to very small values of
this BR of $O(10^{-7})$ or so, whilst the $\left( \delta^{d}_{13}\right)_{LR}$
insertion allows for BR($b \to d \gamma$) ranging from few times $10^{-4}$ up 
to few times $10^{-3}$ for decreasing values of $x=m^{2}_{\tilde{g}}/
m^{2}_{\tilde{q}}$.
 In the SM we expect BR($b \to d \gamma$) to be typically $10-20$
times smaller than BR($b \to s \gamma$), i.e. BR($b \to d \gamma)=(1.7\pm 0.85
)\times 10^{-5}$. Hence a large enhancement in the SUSY case is conceivable if
$\left( \delta^{d}_{13}\right)_{LR}$ is in the $10^{-2}$ range. Notice that in
the MSSM we expect $\left( \delta^{d}_{13}\right)_{LR}<m^{2}_{b}/
m^{2}_{\tilde{q}}\times V_{td}<10^{-6}$, hence with no hope at all of a
sizeable contribution to $b \to d \gamma$.

 \begin{table}
 \begin{center}
 \begin{tabular}{||c|c|c||}  \hline \hline
  & & \\
 $x$ & $\left|\left(\delta^{d}_{23} \right)_{LL}\right| $ &
 $\left|  \left(\delta^{d}_{23} \right)_{LR}\right| $ \\
  & & \\ \hline
 $
   0.3
 $ &
 $
4.4
 $ & $
1.3\times 10^{-2}
 $ \\
 $
   1.0
 $ &
 $
8.2
 $ & $
1.6\times 10^{-2}
 $ \\
 $
   4.0
 $ &
 $
26
 $ & $
3.0\times 10^{-2}
 $ \\ \hline \hline
 \end{tabular}
 \caption[]{Limits on the $\left| \delta_{23}^{d}\right|$ from
 $b\rightarrow s \gamma$ decay for an average squark mass
 $m_{\tilde{q}}=500\mbox{GeV}$  
 and for different values of $x=m_{\tilde{g}}^2/m_{\tilde{q}}^2$. For
 different values of $m_{\tilde{q}}$,  
 the limits can be obtained multiplying the ones in the table by 
 $\left(m_{\tilde{q}}(\mbox{GeV})/500\right)^2$.}
 \label{tab:bsg}
 \end{center}
 \end{table}

An analysis similar to the one of $b \to s \gamma$ decays can be
performed in the leptonic sector where the masses 
$m_{\tilde{q}}$ and $m_{\tilde{g}}$ are replaced by the average slepton
mass $m_{\tilde{l}}$ and the photino mass $m_{\tilde{\gamma}}$ respectively.
In table~\ref{lep} we exhibit the bounds on $\left(\delta^l_{ij}\right)_{LL}$ 
and $\left(\delta^l_{ij}\right)_{LR}$ coming from the limits on 
$\mu\rightarrow e\gamma,~\tau\rightarrow e\gamma$ and 
$\tau\rightarrow \mu\gamma$, for a slepton mass of O(100 GeV)
and for different values of $x=m_{\tilde{\gamma}}^2/m_{\tilde{l}}^2$
\cite{GGMS}. 

 \begin{table}
 \begin{center}
 \begin{tabular}{||c|c|c||}  \hline \hline
  & & \\ 
 $x$ & $\left|\left(\delta^{l}_{12} \right)_{LL}\right| $ &
 $\left|  \left(\delta^{l}_{12} \right)_{LR}\right| $ \\ 
  & & \\ \hline
 $
   0.3
 $ &
 $
4.1\times 10^{-3}
 $ & $
1.4\times 10^{-6}
 $ \\
 $
   1.0
 $ &
 $
7.7\times 10^{-3}
 $ & $
1.7\times 10^{-6}
 $ \\
 $
   5.0
 $ &
 $
3.2\times 10^{-2}
 $ & $
3.8\times 10^{-6}
 $ \\ \hline \hline
  & & \\ 
 $x$ & $\left|\left(\delta^{l}_{13} \right)_{LL}\right| $ &
 $\left|  \left(\delta^{l}_{13} \right)_{LR}\right| $ \\ 
  & & \\ \hline
 $
   0.3
 $ &
 $
15
 $ & $
8.9\times 10^{-2}
 $ \\
 $
   1.0
 $ &
 $
29
 $ & $
1.1\times 10^{-1}
 $ \\
 $
   5.0
 $ &
 $
1.2\times 10^{2}
 $ & $
2.4\times 10^{-1}
 $ \\ \hline \hline
  & & \\ 
 $x$ & $\left|\left(\delta^{l}_{23} \right)_{LL}\right| $ &
 $\left|  \left(\delta^{l}_{23} \right)_{LR}\right| $ \\
  & & \\ \hline
 $
   0.3
 $ &
 $
2.8
 $ & $
1.7\times 10^{-2}
 $ \\
 $
   1.0
 $ &
 $
5.3
 $ & $
2.0\times 10^{-2}
 $ \\
 $
   5.0
 $ &
 $
22
 $ & $
4.4\times 10^{-2}
 $ \\ \hline \hline
 \end{tabular}
 \caption[]{Limits on the $\left| \delta_{ij}^{d}\right|$ from
 $l_j\rightarrow l_i \gamma$ decays for 
 an average slepton mass $m_{\tilde{l}}=100\mbox{GeV}$ and for different values of 
 $x=m_{\tilde{\gamma}}^2/m_{\tilde{l}}^2$. 
For different values of $m_{\tilde{l}}$, 
 the limits can be obtained multiplying the ones in the table by 
 $\left(m_{\tilde{l}}(\mbox{GeV})/100\right)^2$.}
 \label{lep}
 \end{center}
 \end{table}

\section{CP Violation in Low Energy SUSY - First Two Generations}
\label{sec:genCP}

\begin{figure}   
    \begin{center}
    \epsfysize=10truecm
    \leavevmode\epsffile{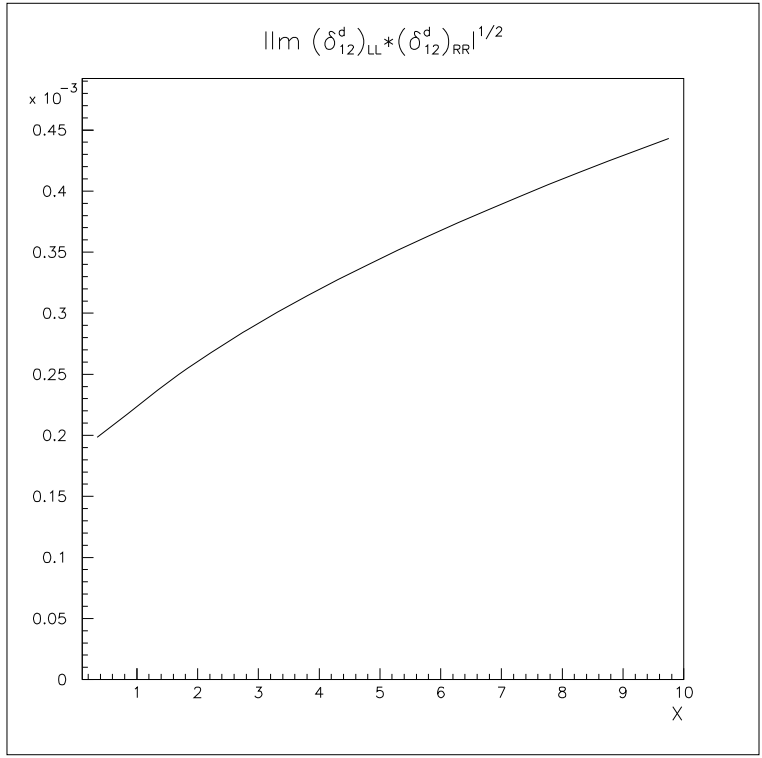}
    \end{center}
    \caption[]{The $\sqrt{\left|{\rm Im}  \left(\delta^{d}_{12}
     \right)_{LL}\left(\delta^{d}_{12}
     \right)_{RR}\right|}$ as a function
     of $x=m_{\tilde{g}}^2/m_{\tilde{q}}^2$, for  an average squark mass
     $m_{\tilde{q}}=500{\rm GeV}$.}
\label{fig:epllrr}
\end{figure}

We start by considering CP violation in the kaon system, i.e. $\varepsilon$ and
$\varepsilon^\prime$. In reference~\cite{GGMS} we provide a very detailed
description of how to calculate the effective Hamiltonian for $\Delta s=2$ and
$\Delta s=1$ processes as well as a determination of the hadronic matrix
elements. Asking for each $\tilde g$ exchange contribution not to exceed the
experimental value $\varepsilon = 2.268 \times 10^{-3}$ we obtain the
bounds reported in table \ref{tab:imds2} for $m_{\tilde q}=500$ GeV.
\begin{table}
 \begin{center}
 \begin{tabular}{|c|c|c|c|}  \hline
  $x$ &
 ${\scriptstyle\sqrt{\left|\Im  \left(\delta^{d}_{12} \right)_{LL}^{2}
\right|} }$ &
 ${\scriptstyle\sqrt{\left|\Im  \left(\delta^{d}_{12} \right)_{LR}^{2}
\right|} }$ &
 ${\scriptstyle\sqrt{\left|\Im  \left(\delta^{d}_{12}
\right)_{LL}\left(\delta^{d}_{12}
 \right)_{RR}\right|} }$ \\
 \hline
 $
   0.3
 $ &
 $
1.5\times 10^{-3}
 $ & $
6.3\times 10^{-4}
 $ & $
2.0\times 10^{-4}
 $ \\
 $
   1.0
 $ &
 $
3.2\times 10^{-3}
 $ & $
3.5\times 10^{-4}
 $ & $
2.2\times 10^{-4}
 $ \\
 $
   4.0
 $ &
 $
7.5\times 10^{-3}
 $ & $
4.2\times 10^{-4}
 $ & $
3.2\times 10^{-4}
 $ \\ \hline
 \end{tabular}
 \caption[]{Limits on
 $\mbox{Im}\left(\delta_{12}^{d}\right)_{AB}\left(\delta_{12}^{d}\right)_{CD}$,
 with $A,B,C,D=(L,R)$, for
 an average squark mass $m_{\tilde{q}}=500\mbox{GeV}$ and for different values
 of
 $x=m_{\tilde{g}}^2/m_{\tilde{q}}^2$. For different values of $m_{\tilde{q}}$,
 the limits can be obtained multiplying the ones in the table by
 $m_{\tilde{q}}(\mbox{GeV})/500$.}
 \label{tab:imds2}
 \end{center}
 \end{table}
In fig.~\ref{fig:epllrr} we plot the bound on $\sqrt{\left|{\rm Im}
\left(\delta^{d}_{12}
\right)_{LL}\left(\delta^{d}_{12}
\right)_{RR}\right|}$ as a function
of $x$ for $m_{\tilde{q}}=500{\rm GeV}$. It should be noticed that the bounds
derived from $\varepsilon$ on the imaginary parts of products of $\delta$'s are
one order of magnitude more stringent than the corresponding limits on the real
parts which are obtained from $\Delta m_K$.

\begin{figure}   
    \begin{center}
    \epsfysize=10truecm
    \leavevmode\epsffile{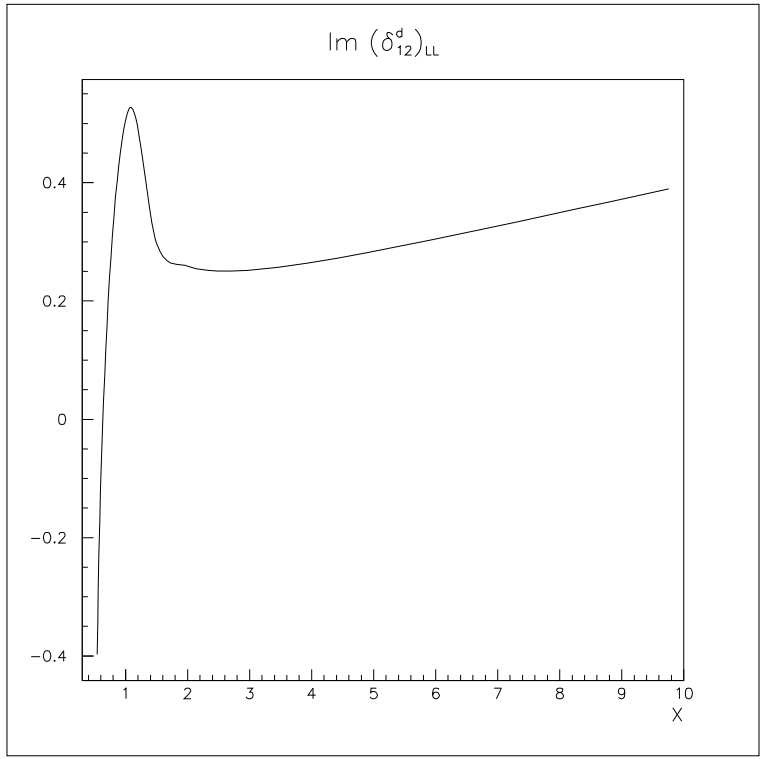}
    \end{center}
    \caption[]{The ${\rm Im}\left(\delta^d_{12}\right)_{LL}$ as a function
     of $x=m_{\tilde{g}}^2/m_{\tilde{q}}^2$, for  an average squark mass
     $m_{\tilde{q}}=500{\rm GeV}$.}
     \label{fig:eppll}
\end{figure}

\begin{figure}   
    \begin{center}
    \epsfysize=10truecm
    \leavevmode\epsffile{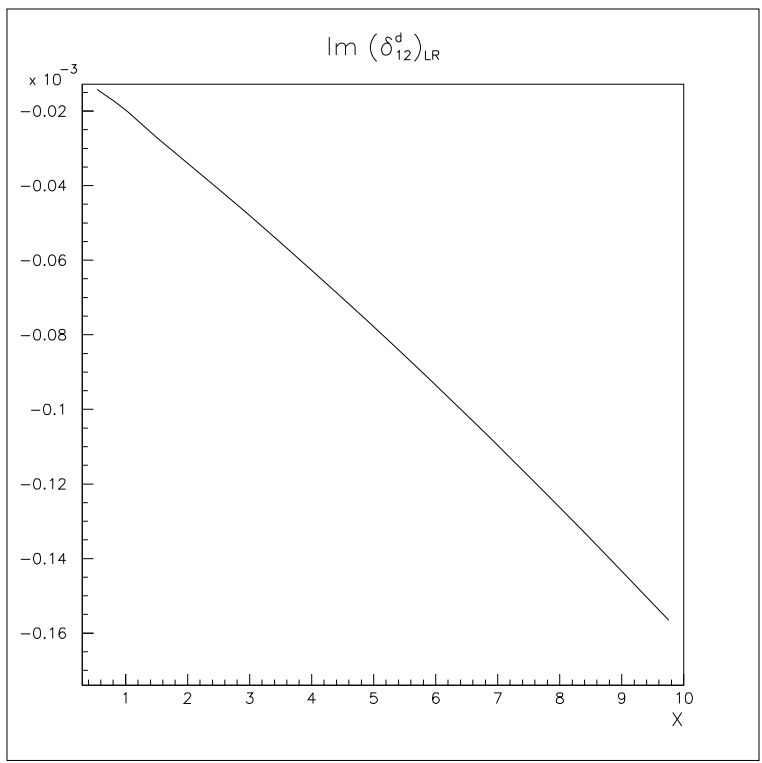}
    \end{center}
    \caption[]{The  ${\rm Im}\left(\delta^d_{12}\right)_{LR}$ as a function
     of $x=m_{\tilde{g}}^2/m_{\tilde{q}}^2$, for  an average squark mass
     $m_{\tilde{q}}=500{\rm GeV}$.}
     \label{fig:epplr}
\end{figure}

Coming to $\Delta s=1$ processes, both superpenguin and superboxes contribute
to $\varepsilon^\prime$. It was only very recently~\cite{GMS} that it was
realized that superboxes are at least as important as superpenguin diagrams in
contributions which proceed through a $\left(\delta^{d}_{12}\right)_{LL}$
insertion. In fig.~\ref{fig:eppll} we report the bound on ${\rm
Im}\left(\delta^d_{12}\right)_{LL}$ as a function of $x$ for $m_{\tilde q}=500$
GeV which comes from the conservative demand that
$\varepsilon^\prime/\varepsilon < 2.7 \times 10^{-3}$. The contribution of box
and penguin diagrams to the LL  terms have opposite signs and a sizeable
cancellation occurs for $x$ close to one, where the two contribution are of
comparable size (this explains the peak around $x=1$ in the plot of
fig.~\ref{fig:eppll}). A much more stringent limit is obtained for
$\left(\delta^{d}_{12}\right)_{LR}$ (fig.~\ref{fig:epplr}). For the LR
contribution only superpenguins play a relevant role. Speaking of
superpenguins, it is interesting to notice that, differently from the SM case,
the SUSY contributions are negligibly affected by electroweak penguins,
i.e. gluino-mediated $Z^0$- or $\gamma$-penguins are strongly suppressed with
respect to gluino-mediated gluon penguins \cite{GGMS}.


In table \ref{tab:imds1} we summarize the bounds on ${\rm Im}
\left(\delta^{d}_{12}  \right)_{LL}$ and ${\rm Im}
\left(\delta^{d}_{12}  \right)_{LR}$ coming from
$\varepsilon^{\prime}/\varepsilon < 2.7 \times 10^{-3}$ for the same values of
SUSY masses chosen in table \ref{tab:imds2}. The comparison of the two tables
leads to the following two conclusions:
\begin{enumerate}
\item if we consider a SUSY extension of the SM where the LR insertions are
much smaller than the LL ones (this is what occurs in the MSSM, for instance),
then fulfilling the bound coming from $\varepsilon$ implies that ${\rm Im}
\left(\delta^{d}_{12}  \right)_{LL}$ is too small to provide a sizeable
contribution to $\varepsilon^{\prime}$ unless $\left(\delta^{d}_{12}
\right)_{LL}$ is almost purely imaginary (remember that $\varepsilon$ bounds
${\rm Im} \left(\delta^{d}_{12}  \right)_{LL}^2$). Hence, in this case the SUSY
contribution would be of superweak nature;
\item if, on the contrary, we have a SUSY model with sizeable LR $\Delta s=1$
mass insertions, then it is possible to respect the bound from $\varepsilon$,
while obtaining a large contribution to $\varepsilon^{\prime}/\varepsilon$. In
this case we would have a SUSY milliweak contribution to CP violation. For this
to occur, we need a SUSY model where $ \left(\delta^{d}_{12}  \right)_{LR}$ is
no longer proportional to $m_s$, but rather to some much larger mass.
\end{enumerate}

 \begin{table}
 \begin{center}
\begin{tabular}{|c|c|c|}   \hline
  & & \\
  $x$ & ${\scriptstyle\left|\Im \left(\delta^{d}_{12}  \right)_{LL}
\right|} $ &
 ${\scriptstyle\left|\Im \left(\delta^{d}_{12}\right)  _{LR}\right| }$
\\
  & & \\\hline
 $
   0.3
 $ &
 $
1.0\times 10^{-1}
 $ & $
1.1\times 10^{-5}
 $ \\
 $
   1.0
 $ &
 $
4.8\times 10^{-1}
 $ & $
2.0\times 10^{-5}
 $ \\
 $
   4.0
 $ &
 $
2.6\times 10^{-1}
 $ & $
6.3\times 10^{-5}
 $ \\ \hline
 \end{tabular}
  \caption[]{Limits from $\varepsilon^{\prime}/\varepsilon < 2.7 \times
  10^{-3}$
 on  $\Im\left(\delta_{12}^{d}\right)$, for
 an average squark mass $m_{\tilde{q}}=500\mbox{GeV}$ and for different values
 of
 $x=m_{\tilde{g}}^2/m_{\tilde{q}}^2$. For different values of $m_{\tilde{q}}$,
 the limits can be obtained multiplying the ones in the table by
 $\left(m_{\tilde{q}}(\mbox{GeV})/500\right)^2$.}
 \label{tab:imds1}
 \end{center}
 \end{table}

The above latter remark would lead us to the natural conclusion that to have
sizeable SUSY contributions to $\varepsilon^{\prime}/\varepsilon$ one needs a
SUSY extension of the SM where $\tilde{q}_L-\tilde{q}_R$ transitions are no
longer proportional to $m_q$ (we remind the reader that in the MSSM a mass term
$\tilde{q}_L\tilde{q}_R^*$ receives two contributions, one proportional to the
parameter A and the other to $\mu$, but both of them are proportional to
$m_q$). However, if this enhancement occurs also for flavour-conserving
$\tilde{q}_L-\tilde{q}_R$ transitions one may envisage some problem with the
very stringent bound on the $d^e_N$. Indeed, imposing this latter constraint
yields the following limits on ${\rm Im}\left(\delta_{11}^{d}\right)_{LR}$ for
$m_{\tilde{q}}=500$ GeV:
\begin{eqnarray}
  x &\qquad&{\rm Im}\left(\delta_{11}^{d}\right)_{LR}\nonumber \\
0.3 & \qquad & 2.4 \times 10^{-6}\nonumber \\
1.0 &  \qquad & 3.0 \times 10^{-6}\nonumber \\
4.0 &  \qquad & 5.6 \times 10^{-6}
  \label{imd11lr}
\end{eqnarray}
A quick comparison of the above numbers with the bounds on ${\rm
Im}\left(\delta_{12}^{d}\right)_{LR}$ from  $\varepsilon^{\prime}/\varepsilon$
reveals that to get a sizeable SUSY contribution to $\varepsilon^{\prime}$ we
need values of  ${\rm
Im}\left(\delta_{12}^{d}\right)_{LR}$ which exceed the bound on ${\rm
Im}\left(\delta_{11}^{d}\right)_{LR}$ arising from $d^e_N$. Obviously, strictly
speaking it is not forbidden for  ${\rm
Im}\left(\delta_{12}^{d}\right)_{LR}$ to be $\ge{\rm
Im}\left(\delta_{11}^{d}\right)_{LR}$, but certainly such a probability does
not look straightforward. In conclusion, although technically it is conceivable
that some SUSY extension may provide a large
$\varepsilon^{\prime}/\varepsilon$, it is rather difficult to imagine how to
reconcile such a large enhancement of  ${\rm
Im}\left(\delta_{12}^{d}\right)_{LR}$ with the very strong constraint on the
flavour-conserving  ${\rm
Im}\left(\delta_{11}^{d}\right)_{LR}$  from $d^e_N$.

\section{CP violation in B Physics}

We now move to the next frontier for testing the unitarity triangle in 
general and in particular CP violation in the SM and its SUSY extensions: $B$ 
physics. We have seen above that the transitions between 1st and 2nd 
generation 
in the down sector put severe constraints on $\Re \delta^d_{12}$ and 
$\Im \delta^d_{12}$  
quantities. To be sure, the bounds derived from $\varepsilon$ and 
$\varepsilon^\prime$ are stronger than the corresponding bounds from $\Delta 
M_K$. If the same pattern repeats itself in the transition between 3rd and 1st 
or 3rd and 2nd generation in the down sector we may expect that the constraints
inferred from $B_d - \bar{B}_d$ oscillations or $b \to s \gamma$ do not prevent
conspicuous new contributions 
 also in 
CP violating processes in $B$ physics. We are going to see below that this is 
indeed the case ad we will argue that measurements of CP asymmetries in 
several $B$-decay channels may allow to disentangle SM and SUSY contributions 
to the CP decay phase.

First, we consider the constraints on $\delta^d_{13}$ and $\delta^d_{23}$ from 
$B_d - \bar{B}_d$ and $b \to s \gamma$, respectively. From the former process 
we obtain the bounds on $\Re \left(\delta^d_{13}\right)^2_{LL}$, $\Re
\left(\delta^d_{13}\right)^2_{LR}$ and  
$\Re \left(\delta^d_{13}\right)_{LL}\left(\delta^d_{13}\right)_{RR}$ which are
reported in table~\ref{reds2}.
The radiative decay $b \to s \gamma$ constraints only the $\left\vert
\left(\delta^d_{23}\right)_{LR}\right\vert$ quantity in a significant way. For
$m_{\tilde q}=500 $ GeV we obtain bounds on $\left\vert
\left(\delta^d_{23}\right)_{LR}\right\vert$  in the range $(1.3 \div 3) \times
10^{-2} $ for $x$ varying from 0.3 to 4, respectively (the bound scales as
$m_{\tilde q}^2$). On the other hand, $b \to s \gamma$ does not limit 
 $\left\vert\left(\delta^d_{23}\right)_{LL}\right\vert$. 
 In the following, we will take 
$\left\vert\left(\delta^d_{23}\right)_{LL}\right\vert=1$ (corresponding to
$x_s=\left(\Delta M/\Gamma\right)_{B_s} > 70 $ for $m_{\tilde q}=500 $ GeV).

New physics can modify the SM predictions on  CP asymmetries in $B$
decays~\cite{rattazzi}
by changing the phase of the $B_{d}$--$\bar{B}_{d}$ mixing
and the  phase and absolute value of  the decay amplitude. 
The general SUSY extension of the SM that we discuss here affects both these 
quantities. 

The remaining part of this chapter tackles the following question of crucial
relevance in the next few years: where and how can one possibly distinguish
SUSY contributions to CP violation in $B$ decays \cite{cptutti}? 
As we said before we want our
answer to be as general as possible, i.e. without any commitment to particular
SUSY models. Obviously, a preliminary condition to properly answer the above
question is to estimate the amount of the uncertainties of the SM predictions
for CP asymmetries in $B$ decays.

To discuss the latter above-mentioned point, 
we choose to work in the theoretical framework of ref.~\cite{cpnoi}.
We use the effective Hamiltonian
(${\cal H}_{eff}$) formalism, including
LO QCD corrections;  in the numerical analysis, we use the LO SM Wilson
coefficients evaluated at $\mu=5$ GeV, as given in ref.~\cite{zeit}. 
In most  of the cases, by choosing different scales (within a reasonable range) 
or by using NLO Wilson coefficients,  the
results vary by about $20-30 \%$.  This is true
with the  exception of some particular channels where uncertainties  are 
larger. The matrix elements of the operators of ${\cal H}_{eff}$
are given in terms of the following Wick contractions 
 between hadronic states: Disconnected
Emission ($DE$), Connected Emission ($CE$), Disconnected Annihilation ($DA$),
Connected Annihilation ($CA$), Disconnected Penguin ($DP$) 
and Connected Penguin
($CP$) (either for left-left ($LL$) or for left-right ($LR$)
current-current operators). Following ref.~\cite{cfms}, where
a detailed discussion can be found, 
instead of  adopting a specific model for  estimating  the different
diagrams,  we let them vary within reasonable 
ranges.  In order to illustrate the relative 
strength and variation of the different contributions, in
table~\ref{tab:ampli}  
we only show, for six different cases,
 results obtained by taking the extreme values of these ranges.
 In the first column 
only $DE=DE_{LL}=DE_{LR}$ are assumed to be different from zero. 
For simplicity, unless stated otherwise,  the same numerical 
values are used for  diagrams corresponding to the insertion 
of  $LL$ or  $LR$ operators, i.e. $DE=DE_{LL}=DE_{LR}$, 
$CE=CE_{LL}=CE_{LR}$, etc.  We then consider, 
in addition to $DE$,  the $CE$ contribution  by taking
 $CE=DE/3$. 
Annihilation diagrams  are included in the third
column, where we use  $DA=0$ and $CA=1/2 DE$ \cite{cfms}. 
Inspired by kaon decays, we allow for some enhancement of 
the matrix elements  of left-right (LR) operators and choose
$DE_{LR}=2 DE_{LL}$ and $CE_{LR}=2 CE_{LL}$  (fourth column). 
Penguin contractions, $CP$ and $DP$, 
can be interpreted  as  long-distance penguin contributions to the matrix 
elements and play an important role:  if   we  
take $CP_{LL}=CE$ and $DP_{LL}=DE$ (fifth column), in some decays 
these terms  dominate the amplitude. 
Finally, in the sixth column,  we allow for long distance
effects which might differentiate penguin contractions with up and charm quarks
in the loop, giving rise to incomplete GIM cancellations (we assume 
$\overline{DP}= DP(c) - DP(u) =  DE/3$ and 
$\overline{CP}= CP(c) - CP(u) =CE/3$). 

\begin{table}
 \begin{center}
 \begin{tabular}{|ccccccc|}
 \hline  
 \ss{Process }&\ss{ $DE$ }&\ss{ $DE+CE$ }&\ss{
 $DE+CE$ }&\ss{ $DE+CE+$ }&\ss{
 $DE+CE+$}&\ss{ $DE+CE+$}\\ \ss{
 }&\ss{   }&\ss{   }&\ss{
 $+CA$ }&\ss{ $CA+LR$ }&\ss{
 $DP+CP$ }&\ss{ $\overline{DP}+ \overline{CP}$}\\
 \hline
&\ss{--}&\ss{--}&\ss{--}& 
\ss{--}&\ss{--}&\ss{--}\\
\ss{
$B^0_d \to J/\psi K_S$}&\ss{-0.03}&\ss{0.1}&
\ss{0.1}&\ss{0.1}&\ss{0.1}&\ss{0.1}\\
 \ss{
}&\ss{-0.008}&\ss{0.02}&\ss{0.02}&
\ss{0.04}&\ss{0.02}&\ss{0.02}\\ 
\hline
&\ss{--}&\ss{--}&\ss{--}&\ss{--}&
\ss{--}&\ss{--}\\ \ss{
$B^0_d \to \phi K_S$}&\ss{0.7}&\ss{0.7}&
\ss{0.7}&\ss{0.6}&\ss{0.4}&\ss{0.4}\\ 
\ss{
}&\ss{0.2}&\ss{0.2}&\ss{0.2}&\ss{0.1}&\ss{0.1}&\ss{0.09}\\
\hline
&\ss{0.08}&\ss{-0.06}&\ss{-0.05}&\ss{-0.02}&\ss{-0.009}&\ss{-0.01}\\ \ss{
$B^0_d \to K_S \pi^0$}&\ss{0.7}&\ss{0.7}&\ss{0.6}&\ss{0.6}&\ss{0.4}&\ss{0.4}\\ 
\ss{
}&\ss{0.2}&\ss{0.2}&\ss{0.2}&\ss{0.1}&\ss{0.1}&\ss{0.09}\\ 
\hline
\ss{$B^0_d \to D^0_{CP} \pi^0$}&\ss{0.02}&\ss{0.02}&\ss{0.02}&\ss{0.02}&
\ss{0.02}&\ss{0.02}\\
\hline
&\ss{-0.6}&\ss{0.9}&\ss{-0.7}&\ss{-2.}&\ss{6.}&\ss{4.}\\ \ss{
$B^0_d \to \pi^0 \pi^0$}&\ss{0.3}&\ss{-0.07}&\ss{0.4}&\ss{-0.4}&\ss{-0.07}&
\ss{-0.06}\\ \ss{
}&\ss{0.06}&\ss{-0.02}&\ss{0.09}&\ss{-0.1}&\ss{-0.02}&\ss{-0.02}\\ 
\hline
&\ss{-0.09}&\ss{-0.1}&\ss{-0.1}&\ss{-0.3}&\ss{-0.9}&\ss{-0.8}\\ \ss{
$B^0_d \to \pi^+\pi^-$}&\ss{0.02}&\ss{0.02}&\ss{0.03}&\ss{0.09}&\ss{0.8}&
\ss{0.4}\\ \ss{
}&\ss{0.005}&\ss{0.006}&\ss{0.008}&\ss{0.02}&\ss{0.2}&\ss{0.1}\\ 
\hline
&\ss{0.03}&\ss{0.04}&\ss{0.05}&\ss{0.1}&\ss{0.3}&\ss{0.2}\\ \ss{
$B^0_d \to D^+D^-$}&\ss{-0.007}&\ss{-0.008}&\ss{-0.01}&\ss{-0.02}&\ss{-0.02}&
\ss{-0.02}\\ \ss{
}&\ss{-0.002}&\ss{-0.002}&\ss{-0.002}&\ss{-0.005}&\ss{-0.006}&\ss{-0.005}\\ 
\hline
&\ss{0}&\ss{0}&\ss{0}&\ss{0}&\ss{0.}&\ss{0.07}\\ \ss{
$B^0_d \to K^0 \bar{K}^0$}&\ss{-0.2}&\ss{-0.2}&\ss{-0.2}&\ss{-0.2}&\ss{-0.09}&
\ss{-0.08}\\ \ss{
}&\ss{-0.06}&\ss{-0.05}&\ss{-0.05}&\ss{-0.04}&\ss{-0.02}&\ss{-0.02}\\ 
\hline
&\ss{--}&\ss{--}&\ss{-0.2}&\ss{-0.4}&\ss{--}&\ss{--}\\ \ss{
$B^0_d \to K^+K^-$}&\ss{--}&\ss{--}&\ss{0.04}&\ss{0.1}&\ss{--}&\ss{--}\\ \ss{
}&\ss{--}&\ss{--}&\ss{0.01}&\ss{0.03}&\ss{--}&\ss{--}\\
\hline
&\ss{--}&\ss{--}&\ss{--}&\ss{--}&\ss{--}&\ss{--}\\ \ss{
$B^0_d \to D^0\bar{D}^0$}&\ss{--}&\ss{--}&\ss{-0.01}&\ss{-0.03}&\ss{--}&
\ss{--}\\ \ss{
}&\ss{--}&\ss{--}&\ss{-0.003}&\ss{-0.006}&\ss{--}&\ss{--}\\ 
\hline
&\ss{-0.04}&\ss{0.1}&\ss{0.1}&\ss{0.3}&\ss{0.1}&\ss{0.1}\\ \ss{
$B^0_d \to J/\psi \pi^0$}&\ss{0.007}&\ss{-0.02}&\ss{-0.02}&\ss{-0.03}&
\ss{-0.02}&\ss{-0.02}\\ \ss{
}&\ss{0.002}&\ss{-0.005}&\ss{-0.005}&\ss{-0.008}&\ss{-0.005}&\ss{-0.005}\\ 
\hline
&\ss{--}&\ss{--}&\ss{--}&\ss{--}&\ss{--}&\ss{--}\\ \ss{
$B^0_d \to \phi\pi^0$}&\ss{-0.06}&\ss{-0.1}&\ss{-0.1}&\ss{-0.1}&\ss{-0.1}&
\ss{-0.1}\\ \ss{
}&\ss{-0.01}&\ss{-0.03}&\ss{-0.03}&\ss{-0.03}&\ss{-0.03}&\ss{-0.03}\\ \hline
 \end{tabular}
 \caption[]{Ratios of amplitudes for exclusive $B$ decays. For each channel, 
whenever two terms  with  different CP phases contribute in the SM, 
we  give the ratio $r$ of the two amplitudes.
For each channel,
the second and third lines, where present, contain the ratios of SUSY to SM 
contributions for SUSY masses  of 250 and 500 GeV respectively.}
 \protect\label{tab:ampli}
 \end{center}
 \end{table}

In addition to the ratios of the different SM contributions to the decay
amplitudes given in
table~\ref{tab:ampli}, obtained letting
the matrix elements vary in the broad range defined above, we also give, in
table~\ref{tab:BR}, the branching ratios for the channels of interest to us.
These branching ratios are obtained following the approach of ref.~\cite{bkpi}.
We use QCD sum rules form factors~\cite{qcdsr} to compute the factorizable $DE$
contribution, then fit $CE$ using the available data on $b \to c$ two-body
decays; $CP$ and $DP$ are extracted from the measured $B \to K \pi$ branching
ratios, $CA$ is varied between 0 and 0.5 and $DA$, $\overline{DP}$ and
$\overline{CP}$ are set to zero.  The range of values in table~\ref{tab:BR}
corresponds to the variation of the CKM angles in the presently allowed range
and to the inclusion of the contributions proportional to $CP$ and $DP$
(see ref.~\cite{bkpi} for further details).   

\begin{table}
 \begin{center}
 \begin{tabular}{|cc|}
 \hline 
 Channel&BR $\times 10^{5}$ \\ \hline
 $B \to J/\psi K_{S}$ &$40$\\ \hline
 $B \to \phi K_{S}$ &$0.6-2$\\ \hline 
 $ B \to \pi^{0} K_{S}$ & $0.02 - 0.4$ \\ \hline 
 $ B \to D^{0}_{CP} \pi^{0}$& $16$\\ \hline 
 $B \to D^{+} D^{-}$ & $30-50$\\ \hline 
 $B \to J/\psi \pi^{0}$ & $2$\\ \hline 
 $B \to \phi \pi^{0}$ &$1-4 \times 10^{-4}$ \\ \hline 
 $B \to K^{0} \bar{K}^{0}$ &$0.007-0.3$\\ \hline 
 $B \to \pi^{+} \pi^{-}$ &$0.2-2$\\ \hline 
 $B \to \pi^{0} \pi^{0}$ & $0.003-0.09$\\ \hline 
 $B \to K^{+} K^{-}$ & $< 0.5 $\\ \hline 
 $B \to D^{0} \bar D^{0}$ & $<20$ \\ \hline
 \end{tabular}
 \caption[]{Branching ratios for $B$ decays.}
 \label{tab:BR}
 \end{center}
 \end{table}
 
Coming to the SUSY contributions, we make use of the Wilson coefficients for
the gluino contribution (see eq. (12) of ref. \cite{GGMS}) and parameterize the
matrix elements as we did before for the SM case. We obtain the ratios of the
SUSY to the SM amplitudes as reported in table~\ref{tab:ampli} for $\tilde q$
and $\tilde g$ masses of 250 GeV and 500 GeV (second and third  row, 
respectively). From the table, 
one concludes that the inclusion of the various terms
in the amplitudes, $DE$, $DA$, etc., 
can  modify the ratio $r$ of  SUSY to SM contributions up to one
order of magnitude.

In terms of the decay amplitude $A$, the CP asymmetry reads 
\begin{equation}
{\cal A}(t) = \frac{(1-\vert \lambda\vert^2) \cos (\Delta M_d t )
-2 {\rm Im} \lambda \sin (\Delta M_d t )}{1+\vert \lambda\vert^2} 
\label{eq:asy}
\end{equation}
with $\lambda=e^{-2i\phi^M}\bar{A}/A$. 
In order to be able to discuss the results  model-independently,
we have labeled as $\phi^M$ the generic  mixing phase.
The ideal case occurs when  one  decay 
amplitude only appears in (or dominates)
a decay process: the CP violating asymmetry is  then determined by the
 total phase   $\phi^T=\phi^M+\phi^D$, where $\phi^D$  is the weak phase
  of the decay.
This ideal situation is spoiled by the presence of
several interfering amplitudes.
If the ratios $r$ in table~\ref{tab:ampli} are small, then the uncertainty on 
the sine of the CP phase is $<  r $, while if $r$ is O(1)  $\phi^T$
receives, in general,  large corrections.

The results of our analysis are summarized in tables~\ref{tab:BR} and
\ref{tab:results} which 
collect the branching ratios and CP phases  for the relevant $B$ decays of
table~\ref{tab:ampli}. $\Phi^D_{SM}$ denotes the decay phase in the SM; for each
channel, when two amplitudes with different weak phases are present, we
indicate the SM phase of the Penguin (P) and Tree-level (T) decay
amplitudes. The range of variation of $r$ in the SM ($r_{SM}$) is deduced from
table~\ref{tab:ampli}. For  $B \to K_S \pi^{0}$ the
 penguin contributions (with a vanishing phase) dominate over the
tree-level amplitude   because the latter is Cabibbo suppressed. 
For the channel $b
\to s \bar s d$  only penguin operators or penguin contractions of
current-current operators  contribute. The phase $\gamma$ is present in the
penguin contractions of the $(\bar b u)(\bar u d)$ operator, 
denoted as $u$-P $\gamma$
in table~\ref{tab:results} \cite{Fleischer}.  
 $\bar b d \to \bar q q $ indicates processes occurring via annihilation 
 diagrams which can be measured
 from the last two channels of table~\ref{tab:results}.
In the case $B \to K^{+} K^{-}$ both
current-current and penguin operators contribute. In $B \to D^{0} \bar
D^{0}$ the contributions
from  the $(\bar b u) (\bar u d)$ and the   $(\bar b c) (\bar c d)$
current-current operators   (proportional to the phase $\gamma$) 
tend to cancel out.

\begin{table}
 \begin{center}
 \begin{tabular}{|ccccccc|}
 \hline 
 \ss{Incl. }&\ss{ Excl. }&\ss{ $\phi^{D}_{\rm SM}$ }&\ss{ $r_{\rm SM}$ }&\ss{ 
 $\phi^{D}_{\rm SUSY}$ }&\ss{ $r_{250}$ }&\ss{ $r_{500}$ }\\ 
 \hline
\ss{ $b \to c \bar c s$ }&\ss{ $B \to J/\psi K_{S}$ }&\ss{ 0 }&\ss{ -- }&\ss{
 $\phi_{23}$ }&\ss{ $0.03-0.1$ 
 }&\ss{$0.008-0.04$ }\\ \hline
 \ss{ $b \to s \bar s s$ }&\ss{ $B \to \phi K_{S}$ }&\ss{ 0 }&\ss{ -- }&\ss{
 $\phi_{23}$ }&\ss{ $0.4-0.7$ }&\ss{ 
 $0.09-0.2$ }\\ \hline \ss{
 $b \to u \bar u s$ }&\ss{ }&\ss{ P $0$ }&\ss{  }&\ss{  }&\ss{  }&\ss{
  }\\ 
\ss{}&\ss{$ B \to \pi^{0} K_{S}$} &\ss{  }&\ss{ $0.01-0.08$ }&\ss{
  $\phi_{23}$ }&\ss{ $0.4-0.7$ }&\ss{ 
 $0.09-0.2$ }\\ \ss{ 
 $b \to d \bar d s$ }&\ss{ }&\ss{ T $\gamma$ }&\ss{  }&\ss{  }&\ss{  }&\ss{
  }\\ \hline \ss{
 $b \to c \bar u d$ }&\ss{ }&\ss{ 0 }&\ss{  }&\ss{  }&\ss{  }&\ss{
  }\\ \ss{ 
  }&\ss{$ B \to D^{0}_{CP} \pi^{0}$ }&\ss{  }&\ss{ 0.02 }&\ss{ -- }&
  \ss{ -- }&\ss{
 -- }\\ \ss{  
 $b \to u \bar c d$ }&\ss{ }&\ss{ $\gamma$ }&\ss{  }&\ss{  }&\ss{  }&\ss{
  }\\ \hline \ss{
  }&\ss{ $B \to D^{+} D^{-}$ }&\ss{ T $0$ }&\ss{ $0.03-0.3$ }&\ss{  }&\ss{
  $0.007-0.02$ }&\ss{ 
  $0.002-0.006$ }\\ \ss{ 
  $b \to c \bar c d$}&\ss{ }&\ss{ }&\ss{  }&\ss{ $\phi_{13}$ }&\ss{ }&\ss{
  }\\ \ss{ 
  }&\ss{ $B \to J/\psi \pi^{0}$ }&\ss{ P $\beta$ }&\ss{ $0.04-0.3$ }&\ss{
  }&\ss{ $0.007-0.03$ }&\ss{ 
  $0.002-0.008$ 
  }\\ \hline \ss{
  }&\ss{ $B \to \phi \pi^{0}$ }&\ss{ P $\beta$}&\ss{ -- }&\ss{ }&\ss{
  $0.06-0.1$ }&\ss{ 
  $0.01-0.03$ }\\ \ss{ 
  $b \to s \bar s d$}&\ss{ }&\ss{  } &&\ss{ $\phi_{13}$ }&\ss{ }&\ss{
  }\\ \ss{ 
  }&\ss{ $B \to K^{0} \bar{K}^{0}$ }&\ss{ {\it u}-P
  $\gamma$  
  }&\ss{ $0-0.07$ }&\ss{ }&\ss{ $0.08-0.2$ }&\ss{
  $0.02-0.06$ 
  }\\ \hline \ss{
 $b \to u \bar u d$ }&\ss{ $B \to \pi^{+} \pi^{-}$ }&\ss{ T
 $\gamma$  
 }&\ss{ $0.09-0.9$ }&\ss{ $\phi_{13}$ }&\ss{ $0.02-0.8$ }&\ss{
 $0.005-0.2$ }\\ \ss{ 
 $b \to d \bar d d$ }&\ss{ $B \to \pi^{0} \pi^{0}$ }&\ss{
 P $\beta$ }&\ss{ $0.6-6$  
 }&\ss{ $\phi_{13}$ }&\ss{ $0.06-0.4$ }&\ss{
 $0.02-0.1$ }\\ \hline \ss{
 }&\ss{ $B \to K^{+} K^{-}$ }&\ss{ T $\gamma$ }&\ss{ $0.2-0.4$ }&\ss{ }&\ss{
 $0.04-0.1$ }&\ss{$0.01-0.03$ 
  }\\ \ss{
 $b \bar d \to q \bar q$ }&\ss{ }&\ss{ }&\ss{ }&\ss{ $\phi_{13}$}&
 \ss{ }&\ss{ }\\ \ss{
 }&\ss{ $B \to D^{0} \bar D^{0}$ }&\ss{ P $\beta$ 
 }&\ss{ only $\beta$ }&\ss{  }&\ss{ $0.01-0.03$ }&\ss{$0.003-0.006$ }\\  \hline
 \end{tabular}
  \caption[]{CP phases for $B$ decays. $\phi^{D}_{SM}$
 denotes the decay phase in 
 the SM; T and P denote Tree and Penguin, respectively; for each
 channel, when two amplitudes with different weak phases are present, 
 one is given in the first row, the other in the last one
 and the ratio of the two  in the $r_{SM}$ column. $\phi^{D}_{SUSY}$
 denotes the phase of the SUSY amplitude, and the ratio of the SUSY to SM
 contributions is given in the $r_{250}$ and $r_{500}$ columns for the
 corresponding SUSY masses.}
 \label{tab:results}
\end{center}
 \end{table}

SUSY contributes to the decay amplitudes with  phases 
induced by  $\delta_{13}$ and
$\delta_{23}$ which we denote as $\phi_{13}$ and $\phi_{23}$. The ratios of
$A_{SUSY}/A_{SM}$ for SUSY masses of 250 and 500 GeV as obtained from 
table~\ref{tab:ampli} are reported in the $r_{250}$ and $r_{500}$ columns 
of table~\ref{tab:results}. 

We now draw some conclusions from the results of table~\ref{tab:results}. 
In the SM, the first
six  decays  measure directly the mixing phase $\beta$, up to
corrections which, in most of the cases, are expected to be small. 
These corrections, due to the presence of  two 
amplitudes contributing with different phases,  produce 
 uncertainties of $\sim 10$\% in   $B \to K_S \pi^{0}$,
 and  of $\sim 30$\%  in $B \to D^{+} D^{-}$ and $B \to
J/\psi \pi^{0}$.   In spite
of the uncertainties,  however, there are cases where
 the SUSY contribution gives rise to significant changes. 
 For example, for SUSY masses of O(250) GeV, SUSY corrections  can  shift the
measured value of the sine of the phase in
 $B \to \phi K_S$ and in $B \to K_S \pi^{0}$ decays by an amount of 
 about 70\%.  For these decays  SUSY effects are sizeable even for 
masses of 500 GeV.  In $B \to
J/\psi K_S$  and $B \to \phi \pi^0$ decays, SUSY effects are only  about $10$\%
but SM uncertainties are negligible.  In $B \to K^0 \bar{K}^0$ 
the larger  effect, $\sim 20$\%,   is partially covered by the 
indetermination of 
about $10$\%  already existing in the SM. 
Moreover the rate for this channel is expected to be rather small.
In $B \to D^{+} D^{-}$  and $B \to K^{+} K^{-}$, SUSY effects are 
completely obscured  by the errors in the estimates of the SM amplitudes.
In $B^0\to D^0_{CP}\pi^0$ the asymmetry  is sensitive to the mixing angle 
$\phi_M$ only because the decay amplitude is unaffected by SUSY. 
This result can be used in connection with $B^0 \to K_s \pi^0$, since
a difference in the measure of the phase  is  a manifestation
of SUSY effects.
\par 
Turning to $B \to \pi \pi$ decays, both the uncertainties
in the SM  and  the SUSY contributions are very large. Here we
witness the presence of three independent amplitudes with different phases 
and of comparable size.  The observation of SUSY effects in
the $\pi^{0} \pi^{0}$ case is hopeless. The possibility of 
separating SM and SUSY contributions  by using the isospin
analysis remains an open possibility which deserves further investigation.
For a thorough discussion of the SM uncertainties in $B \to \pi \pi $ see
ref.~\cite{cfms}. 

In conclusion, our analysis shows that measurements of CP asymmetries in
several channels may allow the extraction of the CP mixing phase and
to disentangle  SM and SUSY contributions to the CP decay phase.  
The golden-plated decays in this respect are $B \to \phi K_S$
and $B \to K_S \pi^0$ channels. The size of the SUSY effects is
clearly controlled by the the non-diagonal SUSY mass
insertions $\delta_{ij}$, which for illustration we have assumed to have the
maximal value compatible with the present experimental limits on 
$B^0_d$--$\bar B^0_d$ mixing.

\section{Outlook}

In the past major emphasis was given to the fact that the MSSM succeeded
to pass all the dangerous FCNC and CP tests unscathed. As important as
this point may actually be, we think that what really matters is how much
these rare processes can yield us a clue on the low-energy SUSY
realization and, hence, on the underlying theory which produces it. In
this view the MSSM constitutes an interesting prototype for more general
SUSY extensions of the SM: FCNC and CP provide crucial ``borders'' on the
allowed departures from such prototype. In particular, the issue of the
way one realizes the breaking of SUSY becomes central for the solution of
the flavour problem. 

On a more phenomenological basis, we hope that these two lectures may help
in correcting a rather common misjudgment on the possibility of achieving
experimental hints of the existence of low-energy SUSY. We refer to the
statement that if at LEP II there is no SUSY manifestation, then we have
to wait for LHC, i.e. quite a few years, before having any answer about
the existence of SUSY. We strongly believe that experiments 
dealing with FCNC and CP violating phenomena have a conspicuous
potentiality to give us some hints of new physics, in particular if the
latter is represented by low-energy SUSY. We have stressed that such a
potentiality is better expressed in three classes of FCNC and CP
experiments which promise to give us important improvements well before
the advent of LHC: i) CP violation in B physics (and, to some extent, also
rare FCNC B decays); ii)  measurements of the electric dipole moments of the
neutron and the electron; iii) flavour lepton number violations
($\mu\rightarrow e\gamma$, $\mu-e$ conversion in nuclei). 

Our conviction that this possibility of manifestation of SUSY is not a
wishful thought, but rests on a solid ground is closely linked to a point
that we hope to have stressed enough in these lectures. The constrained
MSSM extension of the SM is a useful SUSY prototype model to perform
quantitative analysis, but it is unlikely to emerge in its ``minimality''
from an underlying effective supergravity. Indeed, if, for some reason, it
should turn out that it is just this constrained MSSM that is realized at
low energy, then, as we have seen, even our efforts to discover SUSY in
the above three classes of indirect tests would be frustrated. In this
sense  FCNC and CP ``measure'' the amount of departure not only from the SM
physics, but also from the constrained version of the MSSM.
   
\section*{Acknowledgements}

We are grateful to our ``FCNC collaborators" M. Ciuchini, E. Franco, F.
Gabbiani, E. Gabrielli and G. Martinelli who contributed to most of
our recent production on the subject which was reported in these
lectures. A.M. thanks the organizers for the stimulating settling in
which the school took place. The work of A.M. was partly supported by
the TMR project ``Beyond the Standard Model" contract number ERBFMRX
CT96 0090.  L.S. acknowledges the partial support of Fondazione Angelo
della Riccia, Firenze, Italy and of the German Bundesministerium
f{\"u}r Bildung and Forschung under contract 06 TM 874 and DFG Project
Li 519/2-2.

\end{document}